\documentstyle[12pt,a4]{article}
\def\d{{\rm d}}
\begin{document}
\begin{flushright}
FHUP-97-1\\
April, 1997 
\end{flushright}
\vspace{1.0cm}
\begin{center}
{\LARGE \bf Collimation of the e$^+$e$^-$-annihilation event}
\end{center}
\vspace{0.5cm}
\begin{center}
{\large K. $\!$Kimura}\\
Department of Physics, Ochanomizu University\\ Otsuka 2, Bunkyo-ku, Tokyo
112, Japan\\
and\\
{\large K. $\!$Tesima\footnote{E-mail address: tesima@fujita-hu.ac.jp}}\\
Department of Physics, Fujita Health University\\ Kutsukake-cho,
Toyoake-shi, Aichi 470-11, Japan
\end{center}
\date{}
\vspace{2.0cm}
\begin{abstract}
The collimation $C$ of a hadronic event in the e$^+$e$^-$-annihilation is
defined as the average of $\cos\theta$, $C=<\cos\theta>$, where $\theta$ is
the angle of each hadron measured from the thrust axis, and the average is
over all the hadrons produced in an event.   It is an infrared-stable
event-shape parameter.  $1-\overline C$, the difference between the unity
and the average collimation at a given energy, is proportional to the
anomalous dimension of the hadron multiplicity at the leading order in
MLLA.  Its next-to-leading order corrections are calculated.
\end{abstract}

\newpage

When an energetic coloured particle (parton) is produced at a high-energy
collision, many hadrons are produced near the direction of the energetic
parton (jet phenomenon).  The shape of a jet provides us significant
information on the dynamics underlying the multiple hadroproduction.

In order to make quantitative predictions by the perturbative QCD, and to
test them against the experiments, the quantity which characterises the
event shape (event-shape parameter) has to be infrared(IR)-stable at
asymptotically high energies: i.e. insensitive to the IR cutoff under which
the QCD interaction becomes non-perturbative.  Otherwise, the quantity
would depend on the non-perturbative hadronisation process, on which our
theoretical knowledge is very limited.

The thrust $T$ in the e$^+$e$^-$-annihilation is an example of the
IR-stable event-shape parameter.  The thrust distribution is known to be
IR-stable even near $T=1$ (collinear limit)\cite{tes1}\cite{cat}.

In this article, we introduce another IR-stable event-shape parameter, the
collimation $C$ of a hadronic event in the e$^+$e$^-$-annihilation.  It is
defined event by event by
\begin{equation}
C\equiv <\cos\theta > = \frac{\sum_i^n\cos\theta_i}{n}
\end{equation}
where $\theta_i$ is the angle of each hadron measured from the thrust axis.
The average is over all the $n$ hadrons in an event.  $C=1$ in the limit
that all the hadrons are in parallel, while $C=1/\,2$ in the limit that the
angular distribution of hadrons is spherically symmetric.

 Let us analyse  $1-\overline C$, the difference between the unity and the
average collimation at a given energy.  It is defined by
\begin{eqnarray}
1-\overline C&=& \overline{1-\cos\theta} \equiv \frac{2\int_{1/2}^1 \d\eta
2(1-\eta)\frac{\d\overline n}{\d\eta}}{\overline n}\\
 \frac{\d\overline n}{\d\eta} &=& \frac{1}{\sigma_0} \frac{\d\sigma_{(1{\rm
PI})}}{\d\eta}\;,\;\;\overline n =
\frac{2}{\sigma_0}\int_{1/2}^1\d\eta\frac{\d\sigma_{(1{\rm
PI})}}{\d\eta}\;,\;\;\eta =\frac{1+\cos\theta}{2} 
\nonumber
\end{eqnarray}
where $\overline n$ is the average hadron multiplicity, and $\d\overline
n/\d\eta$ its angular density.  $\d\sigma_{(1{\rm PI})}/\d\eta$ is the
one-particle-inclusive cross section of a hadron, and $\sigma_0$ the total
hadronic cross section.   Let us call $1-\overline C$ the
"jet-width"\footnote{The jet-width is determined by the average angular
distribution of hadrons at a given energy.  $\overline C$, therefore, is
not exactly identical to the average of the value of the collimation $C$
measured for each hadronic event.}.

At asymptotically high energies, the average multiplicity and its angular
density are determined by the successive soft-gluon emission, and can be
evaluated by the resummation of the associated double-logarithms to all
orders in $\alpha_s$ (the modified leading-log approximation,
MLLA\cite{bas}\cite{dok1}\cite{dok2}\cite{smi}).

As we shall see below, the jet-width is proportional to the anomalous
dimension $\gamma$ of the multiplicity at the leading order in MLLA: 
\begin{equation}
1-\overline C = (2\ln 2)\gamma +({\rm higher\; order}) 
\end{equation}
where $\gamma$ is defined by the dependence of $\overline n$ on the
centre-of-mass energy $W$ of the event  
\begin{equation}
\gamma =\frac{\d \ln\overline n}{\d \ln W^2}\;.
\end{equation}
We take the QCD mass-scale $\Lambda_{QCD}$ as the unit of energy for the
notational simplicity.  

If the local parton-hadron duality (LPHD\cite{dok1}\cite{azi1}) holds for
the particle flow, the modification of the jet-width by the hadronisation
process disappears at high energies. The measurement of the jet-width thus
provides us a test of the theoretical framework of MLLA+LPHD.

The anomalous dimension $\gamma$, which is proportional to
$\sqrt{\alpha_s}$, is the basic quantity in MLLA.  It is the expansion
parameter in the resummed perturbation theory.   $\gamma$ can be measured
directly by comparing the multiplicity data at different energies (with the
use of its definition (4)). The result, however, remains somewhat ambiguous
because of the uncertainty in the systematic errors in different
experiments.  One of the advantages of using (3) is that the value of
$\gamma$ is determined at each experiment.  (The normalisation constant of
$\overline n$, which cannot be determined by the perturbation theory, is
cancelled between the numerator and the denominator on the rhs of (2).)  We
thus have a less ambiguous comparison of the prediction with the
experimental data.

The determination of $\gamma$ by measuring the jet-width $1-\overline C$,
however, is indirect:  The proportionality is only at the leading order. 
It is, therefore, important to evaluate the higher order corrections.

Roughly speaking, the jet-width is the ratio of the multiplicity in the
large-angle region to the total multiplicity.  The multiplicity at large
angles is also interesting because the anglular
ordering\cite{mue1}\cite{azi3}\cite{dok3}, on which the shower Monte Carlo
simulation programs\cite{web}\cite{sjo}are based, needs modification at
large angles (so-called the "dipole corrections"\cite{dok2}\cite{kit}).

The average multiplicity in the e$^+$e$^-$-annihilation is written in terms
of the multiplicity in a gluon-subjet, $M_g(k_{\perp}^2)$\cite{mun}:
\begin{eqnarray}
\overline n&=&\frac{2}{\sigma_0}\int^1_{1/2}\d\eta_1\int\d
z\left.\frac{\d\sigma}{\d\eta_1\d
z}\right|_{0(\alpha_s)}M_g(k_{\perp}^2)\;. \\
k_{\perp} &=& k\sin\theta_1\;,\;\; z = \frac{2k}{W}\; ,\;\; \eta_1 =
\frac{1+\cos\theta_1}{2}\;. \nonumber\\
M_g(k_{\perp}^2)&=& A\left[\ln k_{\perp}^2 \right]^{\gamma_1} \exp
\left[2\gamma_0\sqrt{\ln k_{\perp}^2}\right]\;,\\\nonumber\\
\gamma_0 &=& \sqrt{\frac{C_A}{2\pi b_0}}\;, \;\;\gamma_1
=-\frac{1}{4}-\frac{N_f}{6\pi b_0} \left(1-\frac{C_F}{C_A}\right)\;,\;\;b_0
= \frac{11C_A-2N_f}{12\pi}\;,\nonumber 
\end{eqnarray}
where $C_A=3$ is the gluon colour-charge, $C_F=4/\,3$ the quark
colour-charge, and $N_f$ the number of active quark flavours.  $k$ is the
momentum of the gluon emitted from the quark-antiquark pair
(e$^+$e$^-\rightarrow q\overline{q}+g$), and $\theta_1$ its polar angle
(the direction of the antiquark is chosen as the $-z$-direction). 
$\d\sigma/(\d\eta_1\d z)$ is the one-gluon emission cross section  at
$O(\alpha_s)$:
\begin{equation}
\left.\frac{1}{\sigma_0}\frac{\d\sigma}{\d\eta_1\d z}\right|_{0(\alpha_s)}=
\frac{C_F\alpha_s}{\pi}\frac{2(1-z)(1-z\eta_1)^2 + (1-z)^2\eta_1^2z^2 +
z^2(1-\eta_1)^2}{z\eta_1(1-\eta_1)(1-2z\eta_1 +z^2\eta_1)}\;.
\end{equation}
The normalisation constant $A$ on the rhs of (6) is not determined by the
perturbation theory, and is cancelled between the numerator and the
denominator in the expression for the jet-width (2).  For simplicity, we
put $A=1$ below.   

The integration over the gluon momentum on the rhs of (5) (with (6) and
(7)) gives at the next-to-leading order
\begin{eqnarray}
\overline n&=&\frac{2C_F}{C_A}\left\{
1-\frac{2\gamma_1-1/\,2+B_0\gamma_0^2}{\gamma_0   \sqrt{\ln (
W^2/4)}}\right\}M_g\left(W^2/4\right)\;,\\
B_0 &=& \frac{3}{2}-\ln 2\;.\nonumber
\end{eqnarray}

The jet-width (2) is given at the leading order in MLLA by
\begin{equation}
1-\overline C = \frac{2}{\overline n}\int_{1/2}^1\d\eta_12(1-\eta_1)\int\d
z \frac{1}{\sigma_0} \frac{\d\sigma}{\d\eta_1\d z}M_g(k_{\perp}^2)\;,
\end{equation}

In (9), we identified the direction of the registered hadron to the
direction of the gluon-subjet\cite{smi}.  It was shown in \cite{kit} that
the correction due to the difference between the two directions vanishes at
the next-to-leading order.

In (9), as well as in (5), the evaluation is made for the forward emission
only ($\eta >1/\,2$), and the result is doubled.  Let us assume that the
gluon is less energetic than the antiquark.  (The case that the gluon is
the most energetic among the initial three partons shall be analysed
later.)  Then the condition $\eta >1/\,2$ implies that the antiquark is the
most energetic and that the thrust axis is in the direction of the
antiquark.  The upper bound for the gluon energy-fraction $z=2k/\,W$ is now
\begin{eqnarray}
z &<& \frac{2\eta_1-1}{\eta_1}\quad ({\rm
for}\;\frac{1}{2}<\eta_1<\frac{3}{4}\,)\;,
\nonumber\\
z &<& \frac{1-\sqrt{\,1-\eta_1}}{\eta_1} \quad ({\rm
for}\;\frac{3}{4}<\eta_1<1\,)\;. 
\end{eqnarray}

Evaluation of (9) at the next-to-leading order (with (6), (7) and (10)) gives
\begin{eqnarray}
(9)&=&\frac{4C_F\gamma_0}{C_A\overline n}\left\{\ln
2\left(1-\frac{\gamma_1-1/\,2}{\gamma_0\sqrt{\ln
(W^2/\,12)}}\right)+\frac{B_1\gamma_0}{\sqrt{\ln (W^2/\,12)}}\right\}
\frac{M_g(W^2/\,12)}{\sqrt{\ln (W^2/\,12)}}\;,\\
B_1&=&-\frac{11}{4}-\frac{9}{2}\ln 2+\frac{29}{8}\ln
3+\frac{\pi^2}{12}+\frac{1}{2}{\rm Sp}(1)+\frac{1}{2}{\rm
Sp}\left(\frac{3}{4}\right)-{\rm Sp}\left(\frac{1}{2}\right)-{\rm
Sp}\left(\frac{1}{4}\right)\;,\nonumber\\
{\rm Sp}(x) &\equiv& \int_0^x \frac{\d y}{y}\ln\frac{1}{1-y}\;.\nonumber
\end{eqnarray}
With the use of (8) for $\overline n$, we obtain at the next-to-leading order
\begin{equation}
\overline{1-\cos\theta_1} =2\ln 2\left\{1+ (-\ln
3+B_0)\gamma\right\}\gamma+2B_1\gamma^2\;,
\end{equation}
where $\gamma$, defined by (4), is
\begin{equation}
\gamma = \frac{\gamma_0}{\sqrt{\ln (W^2/\,12)}} +  \frac{\gamma_1}{\ln
(W^2/\,12)}
\end{equation}

So far, we have considered the one-gluon emission (from the initial
quark-antiquark pair), which produces a gluon-subjet to which the
registered hadron belongs. When another hard gluon is emitted, its recoil
may change the direction of the parent quark.  For such additional
hard-gluon emission, its amplitude has an extra factor of $\alpha_s$. 
Because the jet-width is of $O(\sqrt{\alpha_s})$, the $O(\alpha_s)$
contribution is at the next-to-leading order.  The angle of the
gluon-subjet is expressed as the vector-sum of the recoil angle
$\vec{\theta}_2$ of the parent quark and the gluon emission angle
$\vec{\theta}_1$ from the quark:
$\vec{\theta}=\vec{\theta}_1+\vec{\theta}_2$.  The jet-width is thus
\begin{equation}
1-\overline C = \overline{1-\cos\theta}=
\overline{1-\cos\theta_1}+\overline{1-\cos\theta_2}+\overline{(1-\cos\theta_
1)(1-\cos\theta_2)}\;.
\end{equation}
The third term on the rhs of (14) is $O(\gamma^3)$.  At the next-to-leading
order, therefore, we only need to evaluate the first and the second terms
separately.
In terms of the recoil angle $\theta_2$ and the energy fraction
$z_1=2P_1/\,W$ ($P_1$: the energy of the quark),  the one-gluon emission
cross section at $O(\alpha_s)$ reads
\begin{eqnarray}
\left.\frac{\d\sigma}{\d\eta_2\d z_1}\right|_{0(\alpha_s)}&=&
\frac{C_F\alpha_s}{2\pi}\frac{\sigma_0}{1-\eta_2}\left[\frac{1}{1-2z_1\eta_2
+z_1^2\eta_2}\right. \nonumber\\
&\times&\left.\left\{\frac{z_1^2(1-\eta_2)^2}{1-z_1\eta_2}+z_1(1-z_1)(3\eta_
2-1)\right\}+1-z_1\right]\;,\\
\eta_2 &=&\frac{1+\cos\theta_2}{2}\nonumber
\end{eqnarray}
By the use of (15), the average value of $1-\cos\theta_2$ is calculated to be
\begin{eqnarray}
\overline{1-\cos\theta_2} &=& \frac{1}{\sigma_0}\int_{1/2}^1 \d\eta_2
2(1-\eta_2)\int \d z_1 \left.\frac{\d\sigma}{\d\eta_2\d
z_1}\right|_{0(\alpha_s)}=\frac{C_F\alpha_s}{\pi}B_2\;,\\
B_2 &=& \left\{-\frac{9}{2}+\frac{\pi^2}{3}+7\ln \frac{3}{2}+\ln
2\ln\frac{4}{3}-\frac{3}{2}{\rm Sp}(1)+\frac{1}{2}{\rm
Sp}\left(\frac{3}{4}\right)+{\rm
Sp}\left(\frac{1}{2}\right)\right\}\nonumber
\end{eqnarray}
where the integration over the quark momentum is done under the condition
that the antiquark is the most energetic among the three initial partons.

Finally, let us analyse the case that the gluon is the most energetic among
the initial three partons.  The thrust axis is now identical to the
direction of the hard gluon.  The soft-gluon emission parallel to the gluon
would contribute to $\overline n$ at $O(\alpha_s)$, but does not contribute
to the jet-width.  The soft-gluon emission parallel to the (anti)quark, on
the other hand, gives a next-to-leading contribution to the jet-width.  At
this order, therefore, the direction of the registered hadron can be
identified to the direction of the (anti)quark.  In terms of the polar
angle $\theta_3$ of the quark (the gluon is now in the $-z$-direction), the
one-gluon emission cross section at $O(\alpha_s)$ reads 
\begin{eqnarray}
\left.\frac{\d\sigma}{\d\eta_3\d z_1}\right|_{0(\alpha_s)}&=&
\frac{C_F\alpha_s}{2\pi}\frac{\sigma_0}{1-2z_1\eta_3
+z_1^2\eta_3}\left\{\frac{z_1(1-z_1)}{1-z_1\eta_3} +
\frac{1-z_1}{\eta_3}\right. \nonumber\\
&&\qquad\left.-2z_1(1-z_1)+\frac{2z_1(1-\eta_3)}{(1-z_1)\eta_3}\right\}\;,\\
\eta_3 &=&\frac{1+\cos\theta_3}{2}\;.\nonumber
\end{eqnarray}
By the use of (17), the contribution to the jet-width is calculated to be
\begin{eqnarray}
\overline{1-\cos\theta_3} &=& \frac{2}{\sigma_0}\int_{1/2}^1 \d\eta_3
2(1-\eta_3)\int \d z_1 \left.\frac{\d\sigma}{\d\eta_3\d
z_1}\right|_{0(\alpha_s)}= \frac{C_F\alpha_s}{\pi}B_3\nonumber\\
B_3 &=& \left\{-\frac{19}{2}-6\ln 2+\frac{71}{4}\ln 3+21\ln^2 2+2\ln^2
3-16\ln 2\ln 3\right.\nonumber\\
&&\qquad\qquad\qquad\qquad\qquad\qquad\left.-4{\rm Sp}(1)+6{\rm
Sp}\left(\frac{3}{4}\right)-2{\rm Sp}\left(\frac{1}{2}\right)\right\}\;.
\end{eqnarray}

From (12),(16) and (18), we obtain
\begin{eqnarray}
\overline{1-\cos\theta} &=&
\overline{1-\cos\theta_1}+\overline{1-\cos\theta_2}+\overline{1-\cos\theta_3
}\nonumber\\
&=&(2\ln 2)\gamma+\left\{-2\ln 2\ln 3+2(B_0\ln
2+B_1)\right\}\gamma^2+\frac{C_F\alpha_s}{\pi}(B_2+B_3)\;.
\end{eqnarray}

Numerically, the factors in (19) are $2(B_0\ln 2+B_1)\approx -0.086$,
$B_2\approx 0.432$ and $B_3\approx 0.128$.  The origin of the largest part
of the next-to-leading order correction, $-2(\ln 2\ln 3)\gamma^2$ (in the
second term on the rhs of  (19)), is the upper bound of the gluon energy in
evaluating (9).  In fact, as the gluon energy fraction $z$ increases beyond
the limit of (10), the event-shape becomes more collimated (the thrust axis
is now in the direction of the gluon).

The prediction for the jet-width is listed in Table 1.  In the prediction,
(9) and (7) are evaluated by performing the numerical integration, rather
than by the use of the expansion in $\gamma$ (12).  The corresponding value
of the anomalous dimension is given by (13).  The value of (one-loop) QCD
mass scale is chosen to be $\Lambda_{QCD}=0.15GeV$.  The dependence on the
cm energy $W$ (hence on $\Lambda_{QCD}$) is rather weak over wide range of
$W$. This is because it decreases proportionally to $1/\sqrt{\ln W^2}$. 
Indeed, we cannot use the prediction for the determination of the universal
(but scheme-dependent) QCD mass scale (such as $\Lambda_{\overline{MS}}$)
until the next-to-next order correction is calculated.

\newpage
\vspace{2cm}
\noindent{\large Table 1}

The prediction for the jet-width at various energies, and the corresponding
value of the anomalous dimension.
\vspace{0.5cm}

\begin{tabular}{|c|c|c|} \hline
$E_{cm}$ & $1-<C>$ & $\gamma(W^2/\,12)$\\ \hline
\makebox[20mm]{58GeV}&\makebox[25mm]{0.251} &\makebox[25mm]{0.236}\\ \hline 
92GeV & 0.244 & 0.228\\ \hline 
1000GeV & 0.216 & 0.195\\ \hline
\end{tabular}


\newpage

\vspace{2cm}
\begin{thebibliography}{10}

\bibitem{tes1} K. Tesima and M. F. Wade, Z. Phys. 51 (1991) 43; K. Tesima,
The shape of a Jet, Proceedings of the 25th Rencontres de Moriond (Moriond,
March 1990): Hadronic (1990) p.71; Multiple Hadroproduction at very high
energies, Proceedings of the 2nd International Workshop on Physics and
Experiments with Linear e$^+$e$^-$ Colliders (Waicoloa, April 1993), ed. F.
A. Harris et al. (World Scientific, 1993) Vol.II P.683.

\bibitem{cat} S. Catani, G. Turnock, B. R. Webber and L. Trentadue, Phys.
Lett. B263 (1991) 492; Nucl. Phys. B407 (1993) 3.

\bibitem{smi} J.Smith and K.Tesima, Z. Phys. C49 (1990) 591.

\bibitem{bas} A. Basseto, M. Ciafaloni and G. Marchesini, Phys. Rep. C100
(1983) 201.

\bibitem{dok1} Yu. L. Dokshitzer and S. I. Troyan, Proceedings of the XIX
Winter School of the LNPI Vol.I (1984) 144. 

\bibitem{dok2} Comprehensive review is given in; Yu. L. Dokshitzer, V. A.
Khoze and S. I. Troyan, in Perturbative QCD, Vol.5 ed A.H.Mueller (World
Science, Singapore, 1989) p.241;
Yu. L. Dokshitzer, V. A. Khoze, A. H. Mueller, and S. I. Troyan, Basics of
Perturbative QCD (Editions Frontiers, Paris, 1991). 

\bibitem{azi1} Ya. I. Azimov, Yu. L. Dokshitzer, V. A. Khoze and S. I.
Troyan, Z. Phys. C27 (1985) 65. 

\bibitem{mue1} A. H. Mueller, Phys. Lett. 104B (1981) 161; B. I. Ermolayev
and V. S. Fadin, JETP Lett. 33 (1981) 269; A. Bassetto, M. Ciafaloni, G.
Marchesini and A. H. Mueller, Nucl. Phys. B207 (1982) 189. 

\bibitem{azi3} Ya. I. Azimov, Yu. L. Dokshitzer and V. A. Khoze,
Proceedings of the XVII Winter School of the LNPI Vol.I (1982) 162; Ya. I.
Azimov, Yu. L. Dokshitzer, V. A. Khoze and S. I. Troyan, Proceedings of the
XX Winter School of the LNPI Vol.I (1985) 82; 

\bibitem{dok3} Yu. L. Dokshitzer, V. A. Khoze, A. H. Mueller, and S. I.
Troyan, Rev. Mod. Phys. 60 (1988) 373. 

\bibitem{kit} K. Kimura, M. Kitazawa and K. Tesima, Z. Phys. C72 (1996) 271.

\bibitem{web} G. Marchesimi and B. R. Webber, Nucl. Phys. B238 (1984) 1; B.
R. Webber, Nucl. Phys. B238 (1984) 492; B. R. Webber and G. Marchesini,
Nucl. Phys. B310 (1988) 461.

\bibitem{sjo} G. Gustafson, Phys. Lett. B178 (1987) 453; B. Andersson, G.
Gustafson,  Nucl. Phys. B339 (1990)  393; M. Bengtsson, T. Sjostrand, Phys.
Lett. B185 (1987) 435.

\bibitem{mun} T. Munehisa and K. Tesima, Nucl. Phys. B277 (1986) 849; K.
Tesima, Nucl. Phys. B306 (1988) 849 

\end{thebibliography}
\end{document}